# Topological grain boundary segregation transitions

Vivek Devulapalli[1]*, Enze Chen[2], Tobias Brink[1], Timofey Frolov[3]*, Christian H. Liebscher[1,4]*

[1]Max-Planck-Institut für Eisenforschung GmbH; Düsseldorf, 40237, Germany

[2]Department of Materials Science and Engineering, Stanford University, Stanford, CA 94305, USA.

[3]Materials Science Division, Lawrence Livermore National Laboratory, Livermore, CA 94550, USA.

[4]RC FEMS & Faculty of Physics and Astronomy, Ruhr University Bochum, 44801 Bochum, Germany

*Corresponding author. Email: d.vivek07@gmail.com, frolov2@llnl.gov, christian.liebscher@ruhr-uni-bochum.de

Abstract:
Engineering structure of grain boundaries (GBs) by solute segregation is a promising strategy to tailor the properties of polycrystalline materials. Theoretically it has been suggested that solute segregation can trigger phase transitions at GBs offering novel pathways to design interfaces. However, an understanding of their intrinsic atomistic nature is missing. Here, we combine atomic resolution electron microscopy atomistic simulations to discover that iron segregation to GBs in titanium stabilizes icosahedral units ("cages") that form robust building blocks of distinct GB phases. Due to their five-fold symmetry, the Fe cages cluster and assemble into hierarchical GB phases characterised by a different number and arrangement of the constituent icosahedral units. Our advanced GB structure prediction algorithms and atomistic simulations validate the stability of these observed phases and the high excess of Fe at the GB that is accommodated by the phase transitions.

**Introduction**

Properties of polycrystalline materials are strongly influenced by the structure and composition of grain boundaries (GB), which are internal materials interfaces that have recently been found to adopt distinct 2D states (*1–3*). Transitions between different GB phases was suggested to be triggered by many factors, including the segregation of solutes in multicomponent systems (*4–6*). GB segregation in technologically important alloys can have tremendous impact, ranging from deleterious embrittlement (*7*) to beneficial strengthening (*8*), which underscores the importance of obtaining an atomic-level understanding of GB segregation.

Early on, theoretical models of GB segregation proposed that the enriched solute concentration at the GB could induce a GB phase transition when the solubility limit of the boundary is reached (*5*). This idea was further developed into layering segregation transitions, where different layer states characterized by increasing amounts of impurity segregation were suggested as a general phenomenon for GBs (*9–12*). These theories relied on the idea that progressive segregation leads to gradual occupation of a larger number of crystal planes by the solute atoms, but experimental observations are missing (*3, 13, 14*). Organized solute arrangements in the form of a layer have been observed with atomic resolution in twin boundaries with constrained degrees of structural rearrangement (*15*); however, sufficient direct evidence in more general GBs is scant. Moreover, other studies have shown that solute segregation can induce changes of the structural motifs of GBs (*4, 16–19*).

Here, we show by atomic-resolution imaging and atomistic simulations that iron (Fe) segregation to GBs in hexagonal close-packed (hcp) titanium (Ti) induces topological segregation transitions that are associated with an abrupt increase in the Fe solute excess. Iron is accommodated at the GBs in near-regular icosahedral cage structures that are distinct from atomic arrangements of the equilibrium bulk phases in the TiFe system. The geometric frustration of the cages promotes the formation of several distinct GB phases characterised by a different number of clustered cages, which cannot grow into layers or bulk phases since their five-fold symmetry is incompatible with the adjoining crystals. Our results demonstrate how the formation of icosahedral phases can induce novel topological states of GBs that are reminiscent of glass-like or quasicrystalline structures (*20*). By advancing our atomic-level understanding of GB segregation, these observations offer new possibilities to engineer GB topology in polycrystalline alloys, many of which show the propensity for icosahedral-like structures (*21, 22*).

**Results**

*Atomic resolution observations*

We have developed a technique to obtain defined tilt GBs in Ti using thin film deposition (*23*). A global characterization of the thin film microstructure is shown in Fig. S1. Fig. 1A shows a high angle annular dark field (HAADF)-scanning transmission electron microscopy (STEM) image of a $\Sigma 13$ [0001] $\{\bar{7}520\}$ symmetric tilt GB segment in Ti with no detectable solute segregation. A distinct periodicity of the GB structural unit becomes apparent with sub-units being designated as 'A', 'B' and 'C' in Fig. 1A. These units are similar to ones that have been modelled in Ti (*24*) and other ceramic-based materials (*25, 26*). The GB studied here shows a slight deviation from the exact $\Sigma 13$ misorientation, namely 30° as opposed to 27.8°, which leads to the formation of additional GB defects that separate the 'ABC' units (see Fig. 1A).

In previous work, we found that trace amounts of Fe impurities show a strong anisotropic segregation behaviour within the thin films (*23*). Observing a GB segment decorated with Fe as shown in Fig. 1B at atomic resolution clearly demonstrates a complete rearrangement of the atoms from the 'ABC' configuration to cage-like units (hereafter referred to as "cages"). From the elemental contrast provided by HAADF-STEM imaging, it is readily visible that the cage centres are brighter in contrast compared to the surrounding atomic columns that form the cage, as indicated in Fig. 1B. This suggests that the cage-centre is Fe-rich, while the surrounding shell is Ti-rich. We have verified this observation by near-atomic resolution energy dispersive X-ray and electron energy loss spectroscopy in the STEM as shown in Fig. S2. It is interesting to note that the spacing between the cages in the segregated symmetric $\Sigma 13$ GB remains constant at ~0.9 nm, which corresponds to the spacing between two coincident sites in the $\{\bar{7}520\}$ GB and hence the spacing between two consecutive 'ABC' units of the initial GB.

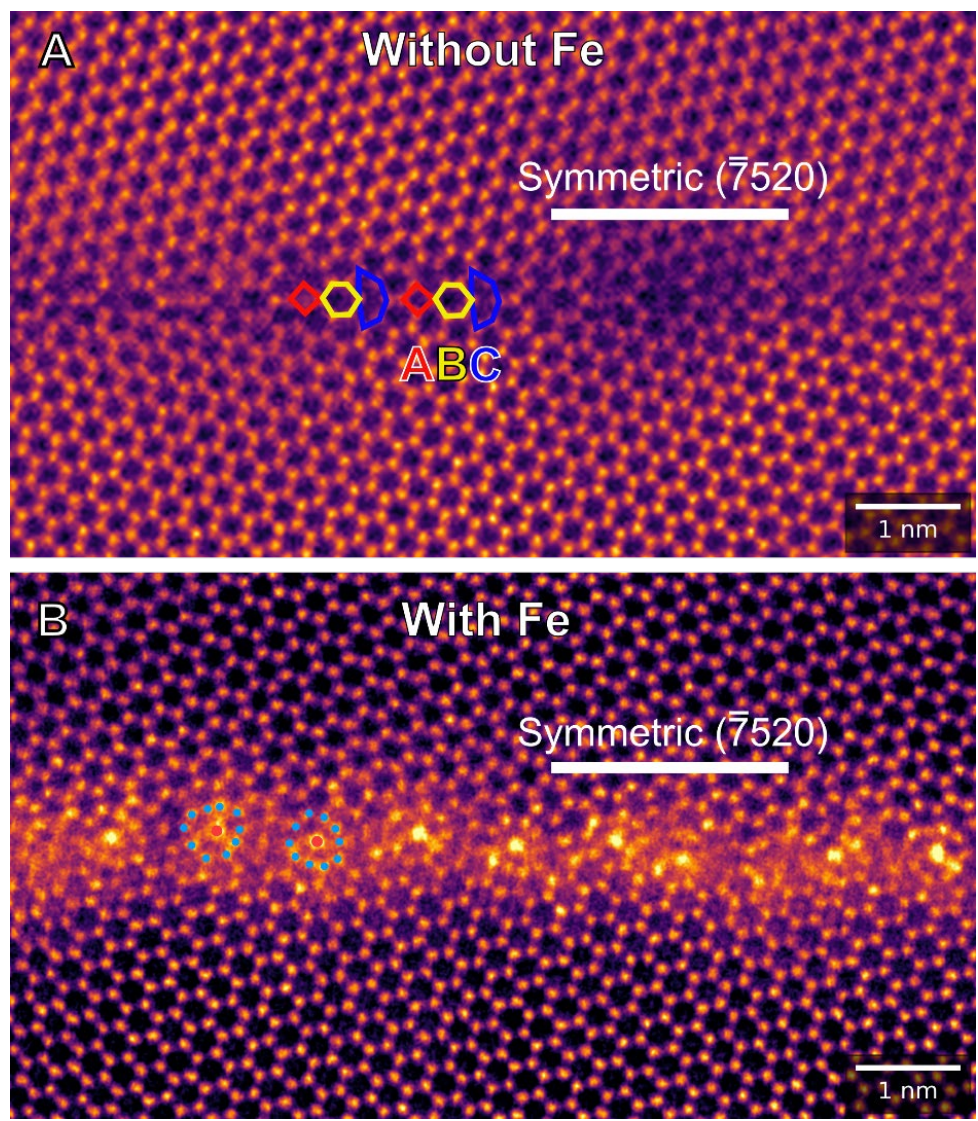

**Fig. 1. Atomic structure of symmetric $\Sigma 13$ [0001] $\{\bar{7}520\}$ tilt GBs in Ti.** (**A**) HAADF-STEM image of the initial GB with no detectable level of segregation demonstrates the periodic arrangement of structural units designated as 'A', 'B' and 'C'. (**B**) Periodic Fe segregation to a GB

resulting in the formation of novel structures at the GB core resembling "cages". The cage centre (red) is rich in Fe, the shell of the cage (blue) rich in Ti, as also verified by atomic scale spectroscopy (see Fig. S2).

We systematically investigated other types of $\Sigma 13$ [0001] GBs with different levels of Fe segregation to decipher the role of Fe-excess on cage density and arrangement. Fig. 2 shows representative atomic resolution images of GBs with increasing cage density. In all images, a similar atomic configuration of the cages is observed to those shown in Fig. 1B. An increased cage spacing of ~2 nm is found for an asymmetric $\Sigma 13$ GB (see Fig. 2A), suggesting an overall lower Fe excess. We want to point out that even isolated cages observed in Fig. 2A are chains of cages stacked on top of each other along the tilt axis of the GB. Interestingly, each of the cages is connected to an 'ABC' structural unit of the initial GB, which is emphasised in Fig. S3. In Fig. 2B, a similar asymmetric GB segment is shown, where complex cage clusters are observed. In certain locations of the GB, even stacking of two or more cages is seen and they seem to arrange themselves in an aperiodic manner within the cluster. Such cage clusters or superstructures are also witnessed at a near-symmetric GB (see Fig. 2C). Here, four cages form a cage cluster, and the clusters themselves constitute an ordered pattern along the GB. In both cases, the increased cage density strongly suggests that the asymmetric GB of Fig. 2B and the near-symmetric GB of Fig. 2C exhibit a higher Fe excess compared to the asymmetric GB shown in Fig. 2A. However, experimental quantification of the GB solute excess is challenging on an atomic scale, particularly using projected HAADF-STEM images of complex 3-dimensional (3D) GB structures. Combining atomic resolution imaging with atomistic simulations provides a way to overcome this limitation and gain deeper insights into the 3D arrangement of atoms within the cage structures (*27*, *28*). To build a correlation between experiment and simulations, we introduce the GB cage density, which is the number of cages per unit length (nm) of the GB. When we compare the different GBs observed in experiment in Fig. 2, we can see an increase in the cage density by a factor of ~4 as shown in Fig. 2D.

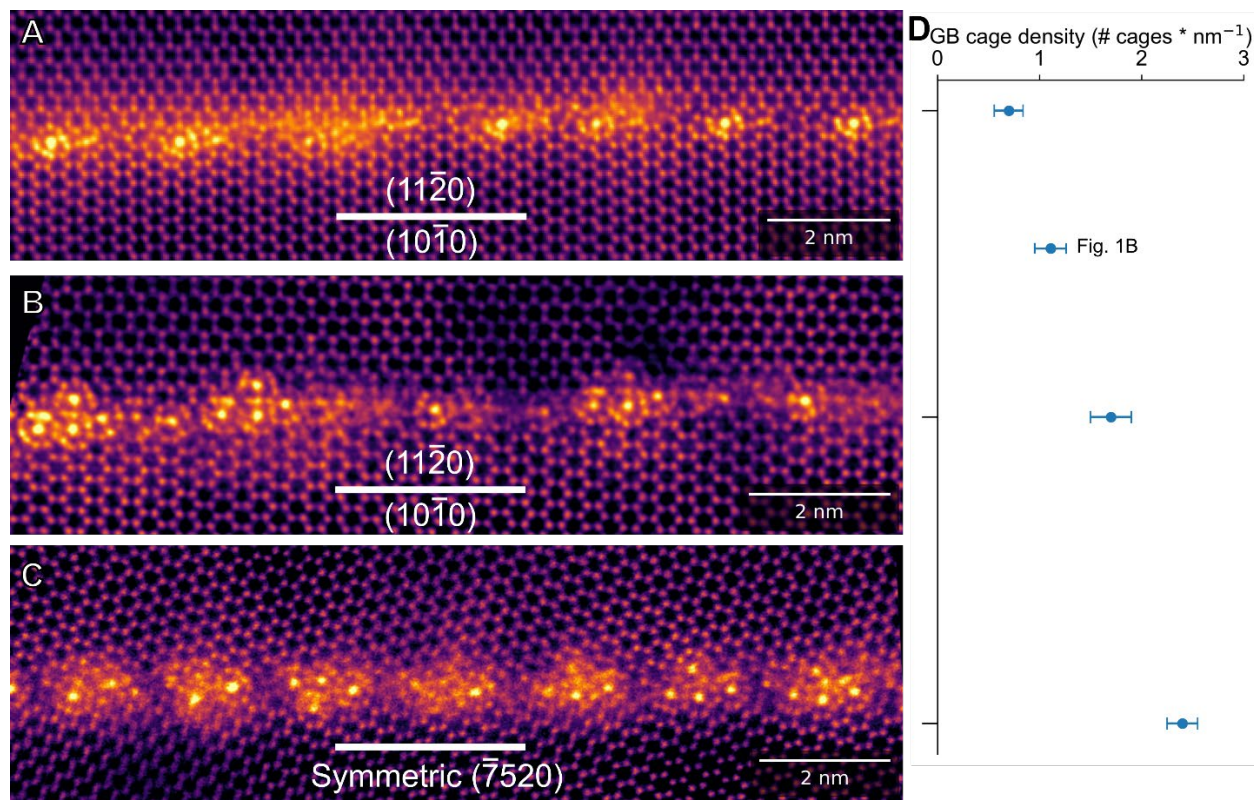

**Fig. 2. Increasing the Fe content results in an increasing GB cage density and cage cluster formation.** (**A**) An asymmetric $\sum 13$ GB with uniformly spaced single cages that are connected to 'ABC' units of the initial GB. (**B-C**) An increasing Fe content in the similar symmetric and asymmetric $\sum 13$ [0001] GBs results in cage clustering. (**B**) Four cages form a distinct arrangement that is repeatedly observed along the GB. (**C**) Four cages assemble to form an aperiodic cluster, which themselves are periodically distributed along the GB. (**D**) GB cage density per nm for the images shown in (A-C). The second data point in (D) corresponds to the symmetric GB with single isolated cages as shown in Fig. 1B.

*Simulations of segregation-induced GB transformations*

We used atomistic simulations to predict the 3D atomic structure of the cage structural units and investigated the nature of the experimentally observed segregation-induced GB phase transitions. Here, we focused on the $\Sigma 13$ $\{\bar{7}520\}$ [0001] GB, which is a representative high-angle, symmetric tilt boundary. We performed GB structure prediction using the Grand Canonical Interface Predictor (GRIP) code as a function of Fe composition at 0K. We also performed finite-temperature canonical molecular dynamics/Monte Carlo (MD/MC) simulations of segregation-induced phase transitions and semi-grand canonical MD/MC simulations to investigate the stability of different GB phases.

Rigorous interface structure prediction requires advanced sampling of possible GB structures. In this work, GB phases of $\Sigma 13$ $\{\bar{7}520\}$ [0001] in Ti and Ti–Fe systems were generated using the GRIP tool. In our previous work, we developed GRIP to study GB phase transitions in elemental metals including Ti. Here, we extended GRIP to multicomponent (binary) systems. The algorithm samples different relative grain translations and densities of lattice sites and optimizes the distribution of chemical components (see Supplementary Methods for details).

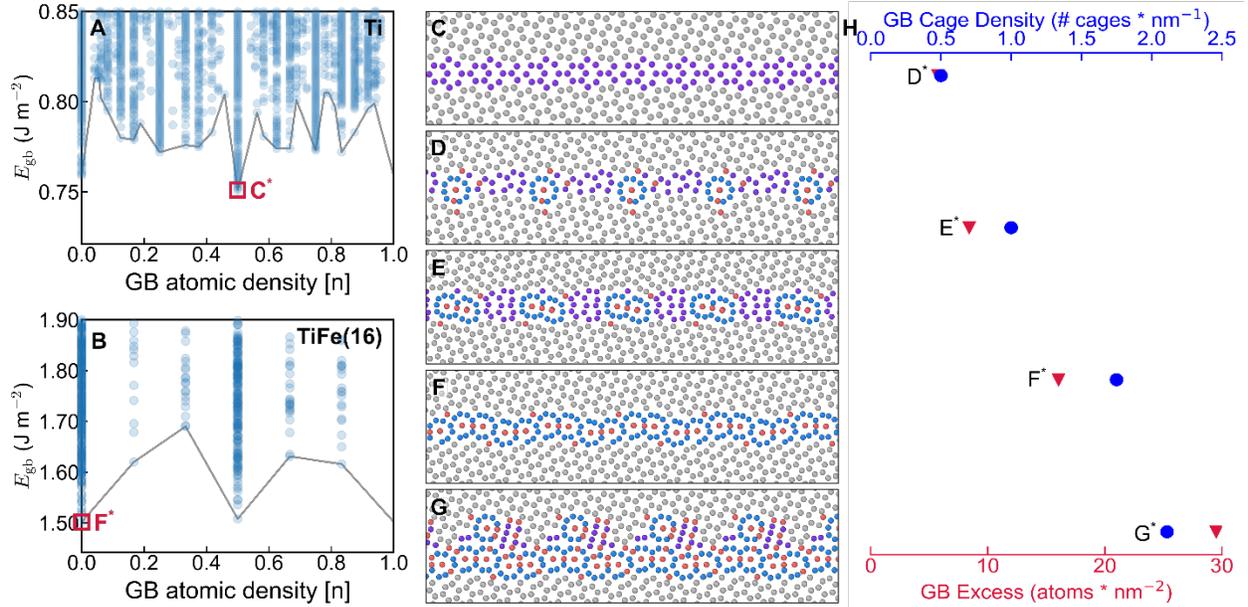

**Fig. 3. GB phases predicted by atomistic simulations for $\Sigma13\{\bar{7}520\}[0001]$ boundary.** (**A-B**) GB structure search using GRIP tool for pure Ti and binary TiFe system with 16 Fe atoms per GB unit cell. GB energy $E_{gb}$ of the generated structures is plotted vs. atomic density [n]. The ground states are indicated by red squares and correspond to [n]=0.5 and [n]=0 for pure and binary cases, respectively, highlighting the coupling between compositional changes and the optimal number of GB lattice sites. (**C-G**) Predicted GB phases as a function of increasing Fe segregation. Fe atoms are shown in red color, all other colors correspond to Ti. (**H**) The GB cage density is increasing with an increase in Fe excess and is in good agreement with the experimentally observed values. The GB solute excess is plotted for each of the four binary states of the GB.

The results of two representative GB structure searches for pure Ti and binary TiFe system with 16 Fe atoms per GB unit cell are shown in Fig. 3A and B, respectively. For pure Ti, the GRIP search shows that the ground-state structure requires grand canonical optimization with the number of extra GB atoms [n]=0.5 of the atomic plane parallel to the boundary plane. On the other hand, for the binary case and 16 atom composition the ground state shown in Fig. 3F is located at [n]=0, indicating the coupling between compositional changes and the optimal number of GB lattice sites. In Ti–Fe, GRIP generated several (5+) distinct GB structures as a function of Fe composition all containing different arrangements of cages as shown in Fig. 3(C-G) and Fig. S8. In this single boundary, the structure search identified the experimentally observed "single-cage," "double-cage," "triple-cage" and "layer of cages" structures, as well as other GB phases that were not found in the experiment shown in Fig. S8. The number of cages at the GB increases to accommodate the increasing Fe segregation.

To study Fe segregation-induced transitions at finite temperature, we performed hybrid MD/MC simulations in the canonical ensemble at $T = 300$K with the total number of atoms of both Fe and Ti kept fixed (details in the Methods). Transformations between the different GB phases shown in Fig. 3 were observed by increasing the amount of Fe in the system in separate simulations. In other words, starting from pure Ti we were able to observe all the different structures in Fig. 3 in our finite temperature simulations. Moreover, prior to each transformation the boundary structure exists in the metastable state for several nanoseconds which is indicative of first-order transitions.

To further confirm that the structures assembled of the different number of cage units can be treated as distinct phases of this boundary and the first-order nature of the phase transitions, we performed additional simulations in the semi-grand canonical ensemble. These simulations revealed the existence of the hysteresis and demonstrated that for the same chemical potential difference $\Delta\mu = \mu_{Fe} - \mu_{Ti}$ and the same concentration of Fe in the bulk, more than one GB structure can be metastable as shown in Fig. 4. By sampling the different $\Delta\mu$ values, we also obtained the limits of stability of each structure. Due to the limitations of the MD/MC algorithm implementation in LAMMPS, we were not able to model large systems and GB phase coexistence to predict the GB phase diagram. The first-order nature of the transformation, however, is further corroborated by the available experimental images in Fig. S4 showing the coexistence of the "initial ABC" and "single-cage" GB structures on the same boundary plane.

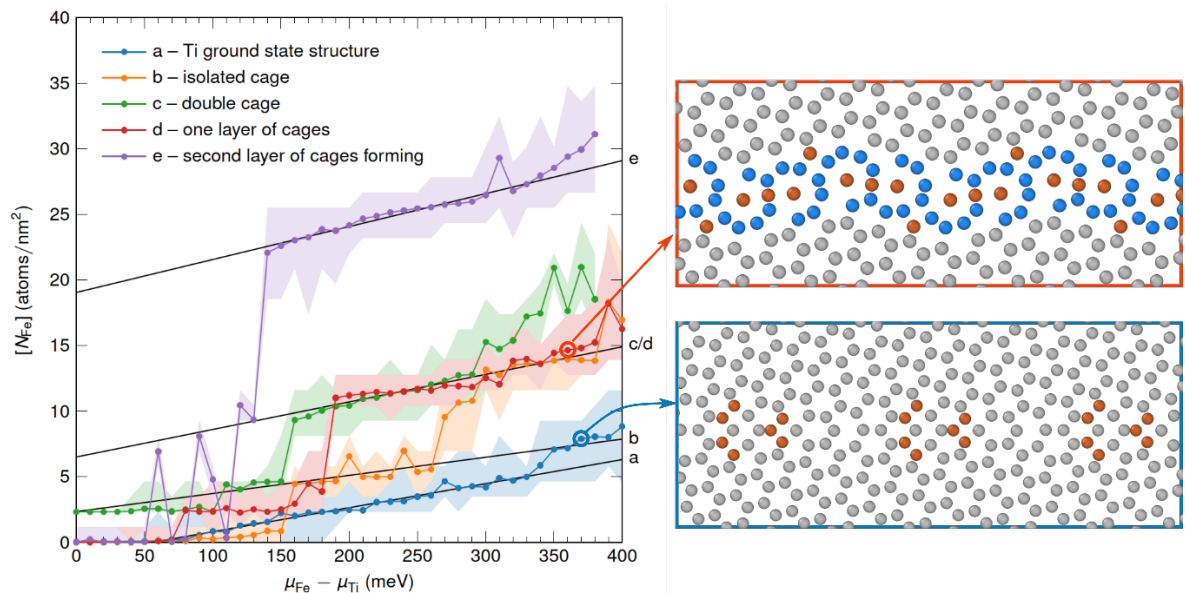

**Fig. 4. GB excess in Fe calculated for different starting GB phases as a function of the chemical potential difference at $T = 300$ K.** The distinct segregation values calculated for different structures at the same chemical potential difference, hysteresis and discontinuous jumps in the segregation values are indicative of the first-order nature of the transitions. Each curve represents MD/MC simulations with a different initial GB structure shown in Fig. 3 (coloured data points). The black lines are guides for the eye that indicate the trends of Fe excess for the different structures. The filled areas indicate the range of excess values encountered during the equilibrated MD/MC simulations.

*Topological structure analysis of Fe-rich cages*
Figure 5 illustrates the three-dimensional atomic structure of the cages predicted by our atomistic simulations. The view of the GB perpendicular to the tilt axis shown in Fig. 5A reveals that the cage units are composed of icosahedra illustrated in Fig. 5B. The structure can be described as a column of Fe atoms occupying interstitial positions in between the (0002) planes in the direction along the tilt axis. The occupancy of Fe in this column is additionally confirmed through STEM simulations, as depicted in Fig. S7. This column is surrounded by a five-fold symmetric ring of Ti atoms within each basal plane, as shown in Fig. 5C. In the two adjacent basal planes, these five-folded rings are rotated relative to each other by 36° to produce an icosahedral cage.

Figure 5D–E show isolated basal planes from the "single-cage" and "double-cage" structures (Fig. 3B and Fig. 3C), respectively. It can be expected that the five-fold symmetry of the Ti rings within each basal plane makes it impossible for cages to cluster and grow indefinitely, because this symmetry is incompatible with periodic crystal structures. This prevents them from forming layered structures across the GB plane or nucleating fully 3D bulk phases. Thus, the symmetry of the cages is likely to be responsible for the unusual appearance of the segregation and structures of the different GB phases in this system. In the "layer-cage" structure where the cages fill the entire plane of the boundary, the Ti rings of cages become distorted and lose their perfect five-fold symmetry.

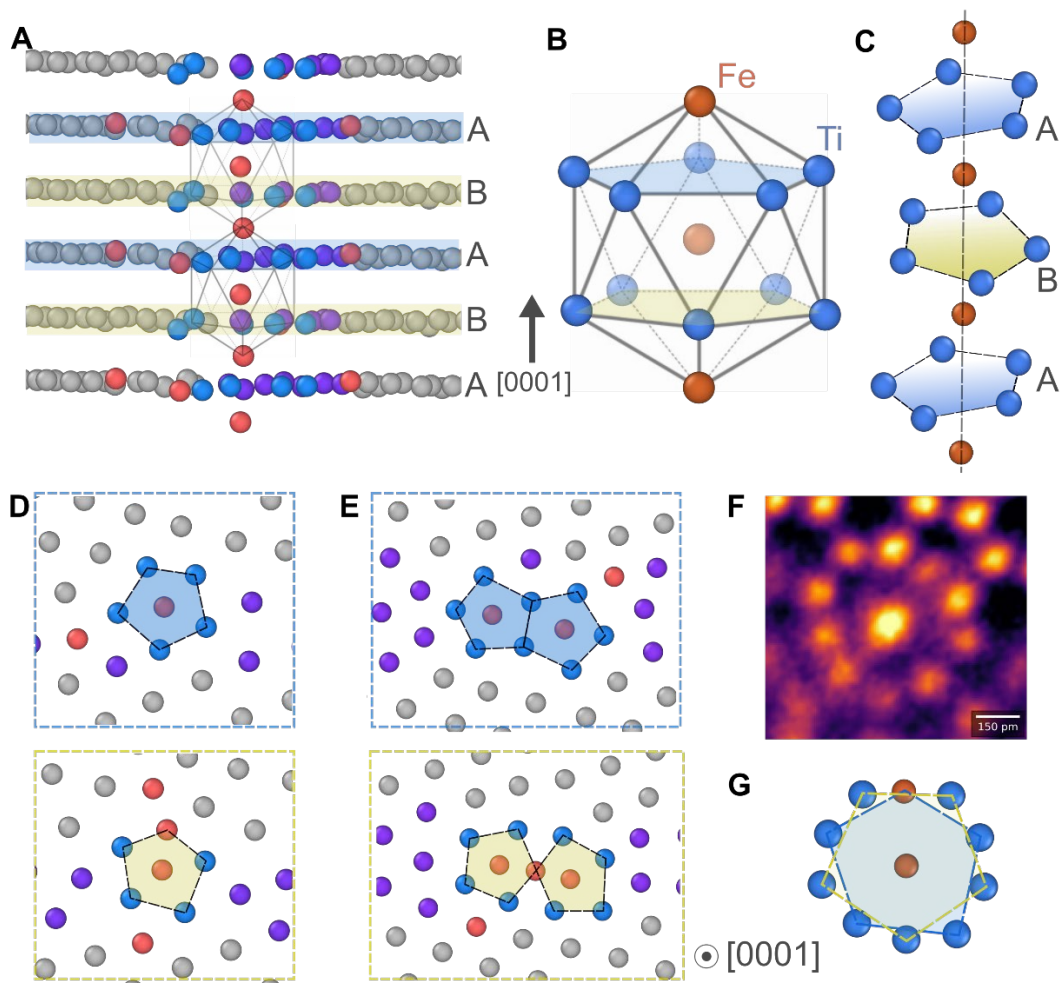

**Fig. 5. 3D structure of Fe cages at GBs.** (**A**) View of a GB containing an isolated cage with the GB plane perpendicular to the plane of figure. The neighbouring basal planes of the adjoining crystals are labelled with A and B. The Fe atoms are shown in red, all other colors represent Ti atoms. Fe (red) segregates to sites in between of every two successive basal planes. (**B**) Structure of the icosahedral 'cage' unit revealing how Fe populates the centre and the opposite vertices of the icosahedron forming (**C**) a column of Fe atoms along the [0001] tilt axis of the GB surrounded by Ti pentagons located in the basal planes. (**D-E**) View of the isolated 'A' and 'B' basal planes with "single-cages" (b) and "double-cage" (c) structures from Fig. 3 revealing the 5-fold symmetry of the cages within each plane and how they attach. (**F**) Magnified view of the cage observed experimentally along the [0001] zone axis where the two pentagons of the icosahedron appear to form a projected 10-atom ring. (**G**) A corresponding cage obtained by atomistic simulations highlighting the differently rotated pentagons (blue and yellow).

**Discussion**

Grain boundary and segregation engineering have been successfully applied to tailor materials properties by interface design. These approaches often assume solute decoration at GBs to occur by classical segregation isotherms without considering structural transitions of the interface and neglect the possibility of novel GB configurations (*29*, *30*). Our direct, atomic-scale observations and atomistic simulations demonstrate that in the Ti–Fe system, Fe segregates to the GBs and creates icosahedral building blocks that represent structurally and compositionally distinct states of the GB. The GBs decorated with these icosahedral units prominently differ from those found in pure hcp-Ti and bcc-Fe. In fact, they are reminiscent of motifs found in C14 Laves phases (here, $TiFe_2$) (*31*), that resemble icosahedra with Fe atoms at the centre and a chemically ordered arrangement of Ti and Fe atoms in the vertices (*32*, *33*). However, the icosahedra observed here primarily contain Ti in the vertices and are geometrically frustrated, leading to the formation of unexpected interface states. Nevertheless, our observations suggest that other bulk systems that show the propensity to form Laves phases or quasicrystals merit further exploration to find novel icosahedral GB phases (*34*, *35*).

The discovered icosahedral segregation motifs are distinct from previous reports of GB segregation phenomena. Existing studies of GB segregation have suggested either progressive layering transitions with increasing solute concentration (e.g., doped alumina (*11*)) or completely distinct GB phases for different concentrations (e.g., Cu–Ag (*4*), Pr-doped ZnO (*25*), Ca-MgO (*17*)). In this study, we combine atomic resolution microscopy and atomistic simulations to observe multiple (5+) distinct GB states in the Ti–Fe system as a function of increasing amounts of Fe; however, the found GB segregation behaviour cannot be classified into either group.

Increasing levels of Fe segregation does not lead to the formation of layers, but rather stabilizes the formation of different configurations of icosahedral cages as building blocks that have five-fold symmetry in each basal plane. The icosahedral segregation motif cannot be described by conventional monolayer- or layering-type transitions, since the topology of the grain boundary core is altered completely. Furthermore, segregation-induced transitions often show only few distinct GB states characterized by different structural units (*4*). Instead, the different GB phases are constructed out of different arrangements of the same icosahedral building blocks. At low Fe concentrations, isolated cages form inside the parent GB structure and some of the 'ABC' units of the clean Ti GB are preserved. At higher Fe concentrations, the cages surprisingly do not form periodic arrays in the boundary plane, instead, two and more cages assemble into cage clusters. The geometric frustration of the cages appears to be the reason for the formation of even more complex aperiodic cage clusters to accommodate higher Fe contents, which, however, cannot grow into periodic layers or even 3D bulk phases (*36*). This unconventional GB segregation behaviour is likely due to the five-fold symmetry of the cages, which is incompatible with crystalline periodicity in two and three dimensions of the adjoining crystals.

The formation of icosahedral cage structures leads to an abrupt increase in the Fe excess concentration of the GB, which is indicative of a first-order GB segregation transition. Hence, the GB can absorb more than two times the amount of Fe atoms compared to the solubility limit of the initial Ti GB structure. Our observed topological segregation transition provides novel insights into solute-induced states of GBs and introduces another lever for interface design. We show that the GB topology can be tailored by the solute excess, which will have important implications on interfacial properties such as GB migration, diffusion (*37*), and corrosion beyond mere solute adsorption.

**Conclusion**

Understanding the structure and phase transitions of GBs in metallic alloys are pivotal to grain boundary structure engineering. Here, we report a segregation-induced phase transition of $\Sigma 13$ [0001] GBs in Ti with distinct, hierarchical phases that depart from conventional theories for GB segregation. Atomically resolved STEM imaging in combination with atomistic simulations are utilized to experimentally observe that the initial GB composed of periodically arranged structural units undergoes a phase transition in the presence of Fe to form icosahedral structures ("cages"). A sudden increase in the GB excess upon cage formation indicates that the GB undergoes a first-order phase transition. The cages act as building blocks for the GB to accommodate Fe segregation. The formation of cages explains how the GB can incorporate more than twice the amount Fe compared to the solubility limit of the initial GB. The GB solute excess determines if the frustrated icosahedrons are spaced evenly along the GB or form clusters. The stability of cages at all Fe concentration levels and even with varying GB planes implies that the observed icosahedral structures are a rather general phenomenon. These observations pave the way forward in utilizing solute segregation as a tool to engineer GBs for a rational design of advanced materials.

**References and Notes**

1.  T. Futazuka, R. Ishikawa, N. Shibata, Y. Ikuhara, Grain boundary structural transformation induced by co-segregation of aliovalent dopants. *Nat Commun* **13**, 5299 (2022).

2.  N. Zhou, T. Hu, J. Luo, Grain boundary complexions in multicomponent alloys: Challenges and opportunities. *Current Opinion in Solid State and Materials Science* **20**, 268–277 (2016).

3.  P. R. Cantwell, M. Tang, S. J. Dillon, J. Luo, G. S. Rohrer, M. P. Harmer, Grain boundary complexions. *Acta Materialia* **62**, 1–48 (2014).

4.  T. Frolov, M. Asta, Y. Mishin, Segregation-induced phase transformations in grain boundaries. *Phys. Rev. B* **92**, 020103 (2015).

5.  M. Guttmann, Grain boundary segregation, two dimensional compound formation, and precipitation. *Metall Trans A* **8**, 1383–1401 (1977).

6.  K. E. Sickafus, S. L. Sass, Grain boundary structural transformations induced by solute segregation. *Acta Metallurgica* **35**, 69–79 (1987).

7.  J. Luo, H. Cheng, K. M. Asl, C. J. Kiely, M. P. Harmer, The Role of a Bilayer Interfacial Phase on Liquid Metal Embrittlement. *Science* **333**, 1730–1733 (2011).

8. T. Yang, Y. L. Zhao, W. P. Li, C. Y. Yu, J. H. Luan, D. Y. Lin, L. Fan, Z. B. Jiao, W. H. Liu, X. J. Liu, J. J. Kai, J. C. Huang, C. T. Liu, Ultrahigh-strength and ductile superlattice alloys with nanoscale disordered interfaces. *Science* **369**, 427–432 (2020).

9. M. Tang, W. C. Carter, R. M. Cannon, Diffuse interface model for structural transitions of grain boundaries. *Phys. Rev. B* **73**, 024102 (2006).

10. M. Tang, W. C. Carter, R. M. Cannon, Grain Boundary Transitions in Binary Alloys. *Phys. Rev. Lett.* **97**, 075502 (2006).

11. S. J. Dillon, M. Tang, W. C. Carter, M. P. Harmer, Complexion: A new concept for kinetic engineering in materials science. *Acta Materialia* **55**, 6208–6218 (2007).

12. J. M. Rickman, J. Luo, Layering transitions at grain boundaries. *Current Opinion in Solid State and Materials Science* **20**, 225–230 (2016).

13. G. S. Rohrer, The role of grain boundary energy in grain boundary complexion transitions. *Current Opinion in Solid State and Materials Science* **20**, 231–239 (2016).

14. S. Yang, N. Zhou, H. Zheng, S. P. Ong, J. Luo, First-Order Interfacial Transformations with a Critical Point: Breaking the Symmetry at a Symmetric Tilt Grain Boundary. *Phys. Rev. Lett.* **120**, 085702 (2018).

15. J. F. Nie, Y. M. Zhu, J. Z. Liu, X. Y. Fang, Periodic Segregation of Solute Atoms in Fully Coherent Twin Boundaries. *Science* **340**, 957–960 (2013).

16. T. Frolov, D. L. Olmsted, M. Asta, Y. Mishin, Structural phase transformations in metallic grain boundaries. *Nat Commun* **4**, 1899 (2013).

17. Y. Yan, M. F. Chisholm, G. Duscher, A. Maiti, S. J. Pennycook, S. T. Pantelides, Impurity-Induced Structural Transformation of a MgO Grain Boundary. *Phys. Rev. Lett.* **81**, 3675–3678 (1998).

18. Z. Wang, M. Saito, K. P. McKenna, L. Gu, S. Tsukimoto, A. L. Shluger, Y. Ikuhara, Atom-resolved imaging of ordered defect superstructures at individual grain boundaries. *Nature* **479**, 380–383 (2011).

19. T. Meiners, T. Frolov, R. E. Rudd, G. Dehm, C. H. Liebscher, Observations of grain-boundary phase transformations in an elemental metal. *Nature* **579**, 375–378 (2020).

20. M. Rappaz, Ph. Jarry, G. Kurtuldu, J. Zollinger, Solidification of Metallic Alloys: Does the Structure of the Liquid Matter? *Metall Mater Trans A* **51**, 2651–2664 (2020).

21. A. K. Sinha, Topologically close-packed structures of transition metal alloys. *Progress in Materials Science* **15**, 81–185 (1972).

22. T. Seki, T. Futazuka, N. Morishige, R. Matsubara, Y. Ikuhara, N. Shibata, Incommensurate grain-boundary atomic structure. *Nat Commun* **14**, 7806 (2023).

23. V. Devulapalli, M. Hans, P. T. Sukumar, J. M. Schneider, G. Dehm, C. H. Liebscher, Microstructure, grain boundary evolution and anisotropic Fe segregation in (0001) textured Ti thin films. *Acta Materialia* **238**, 118180 (2022).

24. Y. C. Wang, H. Q. Ye, On the tilt grain boundaries in hcp Ti with [0001] orientation. *Philosophical Magazine A* **75**, 261–272 (1997).

25. Y. Sato, J.-Y. Roh, Y. Ikuhara, Grain-boundary structural transformation induced by geometry and chemistry. *Phys. Rev. B* **87**, 140101 (2013).

26. K. Inoue, J.-Y. Roh, K. Kawahara, M. Saito, M. Kotani, Y. Ikuhara, Arrangement of polyhedral units for [0001]-symmetrical tilt grain boundaries in zinc oxide. *Acta Materialia* **212**, 116864 (2021).

27. S. Azuma, N. Shibata, S. D. Findlay, T. Mizoguchi, T. Yamamoto, Y. Ikuhara, HAADF-STEM observations of a Σ13 grain boundary in α-Al2O3 from two orthogonal directions. *Philosophical Magazine Letters* **90**, 539–546 (2010).

28. L. Langenohl, T. Brink, G. Richter, G. Dehm, C. H. Liebscher, Atomic-resolution observations of silver segregation in a [111] tilt grain boundary in copper. *Phys. Rev. B* **107**, 134112 (2023).

29. T. Watanabe, An approach to grain boundary design for strong and ductile polycrystals. *Res mech* **11**, 47–84 (1984).

30. D. Raabe, M. Herbig, S. Sandlöbes, Y. Li, D. Tytko, M. Kuzmina, D. Ponge, P.-P. Choi, Grain boundary segregation engineering in metallic alloys: A pathway to the design of interfaces. *Current Opinion in Solid State and Materials Science* **18**, 253–261 (2014).

31. J. L. Murray, The Fe−Ti (Iron-Titanium) system. *Bulletin of Alloy Phase Diagrams* **2**, 320–334 (1981).

32. M. Šlapáková, A. Zendegani, C. H. Liebscher, T. Hickel, J. Neugebauer, T. Hammerschmidt, A. Ormeci, J. Grin, G. Dehm, K. S. Kumar, F. Stein, Atomic scale configuration of planar defects in the Nb-rich *C*14 Laves phase NbFe2. *Acta Materialia* **183**, 362–376 (2020).

33. F. Stein, A. Leineweber, Laves phases: a review of their functional and structural applications and an improved fundamental understanding of stability and properties. *J Mater Sci* **56**, 5321–5427 (2021).

34. K. F. Kelton, P. C. Gibbons, P. N. Sabes, New icosahedral phases in Ti-transition-metal alloys. *Phys Rev B Condens Matter* **38**, 7810–7813 (1988).

35. Z. Yang, L. Zhang, M. F. Chisholm, X. Zhou, H. Ye, S. J. Pennycook, Precipitation of binary quasicrystals along dislocations. *Nat Commun* **9**, 809 (2018).

36. A. Hirata, L. J. Kang, T. Fujita, B. Klumov, K. Matsue, M. Kotani, A. R. Yavari, M. W. Chen, Geometric Frustration of Icosahedron in Metallic Glasses. *Science* **341**, 376–379 (2013).

37. S. V. Divinski, H. Edelhoff, S. Prokofjev, Diffusion and segregation of silver in copper Σ5(310) grain boundary. *Phys. Rev. B* **85**, 144104 (2012).

38. E. Chen, T. W. Heo, B. C. Wood, M. Asta, T. Frolov, Grand canonically optimized grain boundary phases in hexagonal close-packed titanium. arXiv arXiv:2404.04230 [Preprint] (2024). https://doi.org/10.48550/arXiv.2404.04230.

39. I. Sa, B.-J. Lee, Modified embedded-atom method interatomic potentials for the Fe–Nb and Fe–Ti binary systems. *Scripta Materialia* **59**, 595–598 (2008).

40. A. P. Thompson, H. M. Aktulga, R. Berger, D. S. Bolintineanu, W. M. Brown, P. S. Crozier, P. J. in 't Veld, A. Kohlmeyer, S. G. Moore, T. D. Nguyen, R. Shan, M. J. Stevens, J. Tranchida, C. Trott, S. J. Plimpton, LAMMPS - a flexible simulation tool for particle-based materials modeling at the atomic, meso, and continuum scales. *Computer Physics Communications* **271**, 108171 (2022).

41. B. Sadigh, P. Erhart, A. Stukowski, A. Caro, E. Martinez, L. Zepeda-Ruiz, Scalable parallel Monte Carlo algorithm for atomistic simulations of precipitation in alloys. *Phys. Rev. B* **85**, 184203 (2012).

**Acknowledgments:** Authors thank Dr. Marcus Hans for his assistance with deposition of the investigated thin films.

**Funding:** VD and CHL acknowledge funding from the KSB Stiftung. CHL and TF thank the DFG for partial funding through the grant DFG LI 2133/7-1.
This work was performed under the auspices of the U.S. Department of Energy (DOE) by Lawrence Livermore National Laboratory under contract DE-AC52-07NA27344. Prepared by LLNL under Contract DE-AC52-07NA27344. TF was supported by the U.S. DOE, Office of Science under an Office of Fusion Energy Sciences Early Career Award. Computing support for this work came from the Lawrence Livermore National Laboratory Institutional Computing Grand Challenge program.

**Code availability:** The GRand canonical Interface Predictor (GRIP) tool will be accessible at https://github.com/enze-chen/grip upon publication. Other simulation scripts are available from the corresponding authors upon reasonable request.

# Supplementary Materials for

## Topological grain boundary segregation transitions

Vivek Devulapalli*, Enze Chen, Tobias Brink, Timofey Frolov*, Christian H. Liebscher*

Corresponding author: d.vivek07@gmail.com, frolov2@llnl.gov, christian.liebscher@ruhr-uni-bochum.de

**The PDF file includes:**

    Methods
    Figs. S1 to S8

**Methods**

*Thin film preparation*

We described the details of thin film deposition in our earlier work (*23*). Thin films of Ti were deposited onto a SrTiO$_3$ substrate (Crystal GmbH, Germany) using pulsed magnetron sputtering in a commercial system (Ceme Con AG CC 800-9) at MCh, RWTH Aachen, Germany. Ti target of above 99% nominal purity with 0.2 wt. % Fe (0.17 at. %), 0.18 wt. % O (0.54 at. %) and 0.1 wt. %C (0.54 at. %) as a major impurity was deployed. The base pressure before the deposition was 2.2 * 10$^{-6}$ mbar. Sputter power of 1500 W with a pulse duration of t$_{on}$/t$_{off}$ of 200/1800 µs was deployed. This led to ion deposition current (I$_{ion}$) of 40 A, a peak target power density of 46 W cm$^{-2}$ that resulted in a dense film. After the deposition, the film was annealed inside the deposition chamber without breaking the vacuum at the temperature same as the deposition temperature of 600° C for 8h.

*Material Characterisation*

A Thermo Fischer Scientific Scios 2 HiVac dual-beam focused ion beam (FIB)/SEM equipped with EDAX Velocity™ EBSD camera was used to map the grain orientations and identify the coincident site lattice (CSL) GBs. FIB with a Ga+-ion source on the same instrument was used to obtain a plane view (S)TEM lamella. The milling beam current was gradually reduced in multiple steps from 1 nA at 30 kV to 27 pA at 2 kV to achieve a thickness of <100 nm. To conduct HAADF-STEM imaging, STEM-EDS analysis and EELS, a probe corrected Titan Themis 80-300 (Thermo Fischer Scientific) was used. For STEM imaging, the acceleration voltage was set to 300 kV with a probe-semi convergence angle of 23.8 mrad. The inner and outer semi-collection angles of the HAADF detector were set to ~50-200 mrad. A beam current of less than 50 pA was used to reduce electron beam damage. The acceleration voltage was reduced to 120 kV for near-atomic resolution STEM-EDS elemental mapping to increase the inelastic scattering cross-section and further reduce electron beam radiation effects. The core-loss EELS profiles were acquired using Gatan GIF Quantum ERS post-column imaging filter from Ti L$_{2,3}$, O K and Fe L$_{2,3}$ edges using an energy dispersion of 0.25 eV/channel. The raw data and employed Fourier filters (along with the jupyter notebooks) are available as extended data.

*MD/MC simulations*

GB phases of the Σ13{$\bar{7}$520}[0001] symmetric tilt GB in pure Ti and Ti-Fe are generated using the GRand canonical Interface Predictor (GRIP) tool introduced in our earlier work (*38*). Previously, we validated the tool by demonstrating its performance on optimizing GB structures in fcc, bcc, hcp, and diamond cubic systems. Here, we demonstrate the robustness of the approach by extending it to binary systems. The structure sampling using GRIP has three main stages: structure initialization, structure and composition optimization via dynamic sampling, and structure relaxation to 0K. GRIP initiates thousands of optimization routines that are performed in parallel to rigorously sample possible GB degrees of freedom.

During the first stage, the different structures are prepared by randomly sampling relative translations, GB atomic density in pure Ti, and insertion of solute Fe atoms. To increase the diversity of initial configurations, we randomly choose some Fe atoms to substitute for Ti atoms and place the rest in interstitial sites in the GB region. This is necessary as we found that Fe occupies substitutional sites in the perfect hcp lattice while it often occupies interstitial sites in the GB; more generally, the interstitial swaps create a different initial GB structure for the same relative grain translation and atomic density, thus improving the diversity of the initial configurations prior to the dynamic sampling.

During the second stage, the prepared initial configuration evolves via dynamic sampling to predict the equilibrium structure. In our previous study (*38*), we demonstrated that dynamic sampling by standard molecular dynamics (MD) simulations is effective at predicting GB structures in elemental systems. In binary systems, however, this kind of sampling is known to be inefficient because diffusion is too slow to allow for equilibrium redistribution of chemical components. Therefore, we used an established hybrid molecular dynamics/Monte Carlo (MD/MC) simulation to perform the dynamic sampling of the binary system. In this study, we choose the canonical ensemble to only allow swaps between the different atoms (Ti and Fe) and to preserve a constant chemical composition. In the binary system, we cannot calculate GB energy directly the same way we can in an elemental system because the chemical potentials are not known in finite temperature MD/MC and they are not well defined at 0 K. Instead, for the binary searches we calculate what can be described as a GB formation energy which is the difference between the energy of a region containing GB and a bulk hcp region containing the same number of Ti atoms without any Fe. The ground-state structure for a given number of introduced Fe atoms is then identified in the search as the structure with the lowest GB formation energy. This approach only works in cases where all solute atoms go to the GB and the bulk crystal remains essentially pure. This approximation works well for our system because at low temperatures the concentration of Fe in the bulk is extremely small and vanishes at 0 K.

For pure Ti, the GRIP search shows that the ground-state structure requires grand canonical optimization with the number of extra GB atoms $n = 0.5$ of the atomic plane parallel to the boundary plane. The ground state predicted by the modified embedded-atom method (MEAM) interatomic potential used (*39*) had a GB energy of $E_{gb} = 0.751$ J m$^{-2}$. At higher concentrations of Fe, the minimum-energy GB structures optimized by GRIP are shown in Supplementary Fig. S8, with the number of cages increasing with Fe segregation, consistent with experiments. Fig. S9 further demonstrates the robust sampling of GB structures at a fixed concentration of Fe.

To study Fe segregation-induced transitions at finite temperature, we perform hybrid MD/MC simulations. The simulations are performed in the canonical ensemble as implemented in LAMMPS (*40*) at T = 300 K with the total number of both Fe and Ti atoms fixed. During the simulations, some fraction of the Fe atoms segregates to the grain boundary, while some atoms go to the bulk to establish the equilibrium bulk concentration.

Grain boundary structures with a different number of cage units are generated in separate simulations, by gradually increasing the amount of Fe in the system in small increments ranging from 2 to 10 atoms at a time. The new solute atoms are initially inserted randomly everywhere in the system including bulk. Each time, after additional Fe atoms are introduced into the system, we perform simulations for several nanoseconds to either observe transformations of the GB structure which involves nucleation of new cage units or confirm that the initial boundary structure does not change at this concentration. In the latter case, we start a new simulation by increasing the Fe concentration further. We also perform test simulations where a large number of Fe atoms are simultaneously introduced into the pure Ti boundary to confirm that we reach the same state for the given concentration. For both smaller and larger Fe concentrations, the MD/MC simulations are performed using 1 × 1 and 2 × 3 reconstructions, to make sure that the predicted GB structure is not significantly affected by the periodic boundary conditions.

The simulations of even larger area reconstructions quickly become impractical because the MC swaps as implemented in LAMMPS require the evaluation of the energy of the entire system for each swap attempt. As a result, simulations of two-phase coexistence in this model are not feasible. To investigate whether the structures containing different numbers of cage units can be treated as different phases of this boundary, we perform additional MD/MC simulations of each of the obtained structures in the semi-grand canonical ensemble as implemented in LAMMPS. For these, we used $T$ = 300 K with the total number of atoms fixed, but the concentration of both Fe and Ti being controlled by their chemical potential difference, $\Delta\mu = \mu_{Fe} - \mu_{Ti}$ (*41*). We perform 100 MC swaps every 20 MD steps, starting each time from one of the structures obtained in the previous long-time MD simulations and using different values of $\Delta\mu$. The total MD time for each simulation is 5 ns.

These simulations reveal the existence of the hysteresis, meaning that for the same $\Delta\mu$ and the same concentration of Fe in the bulk, more than one GB structure can be metastable. By sampling $\Delta\mu$, we also obtain the limits of stability of each of the structure, where at some $\Delta\mu$ values some structures become unstable and the number of cages either increases or decreases. (The prediction of the $\Delta\mu$ values at which stable coexistence of different phases could occur requires the evaluation of the GB free energy as a function of Fe concentration. Unfortunately, the available data would only allow us to evaluate the change in the free energy and not the absolute value for each of the structures.)

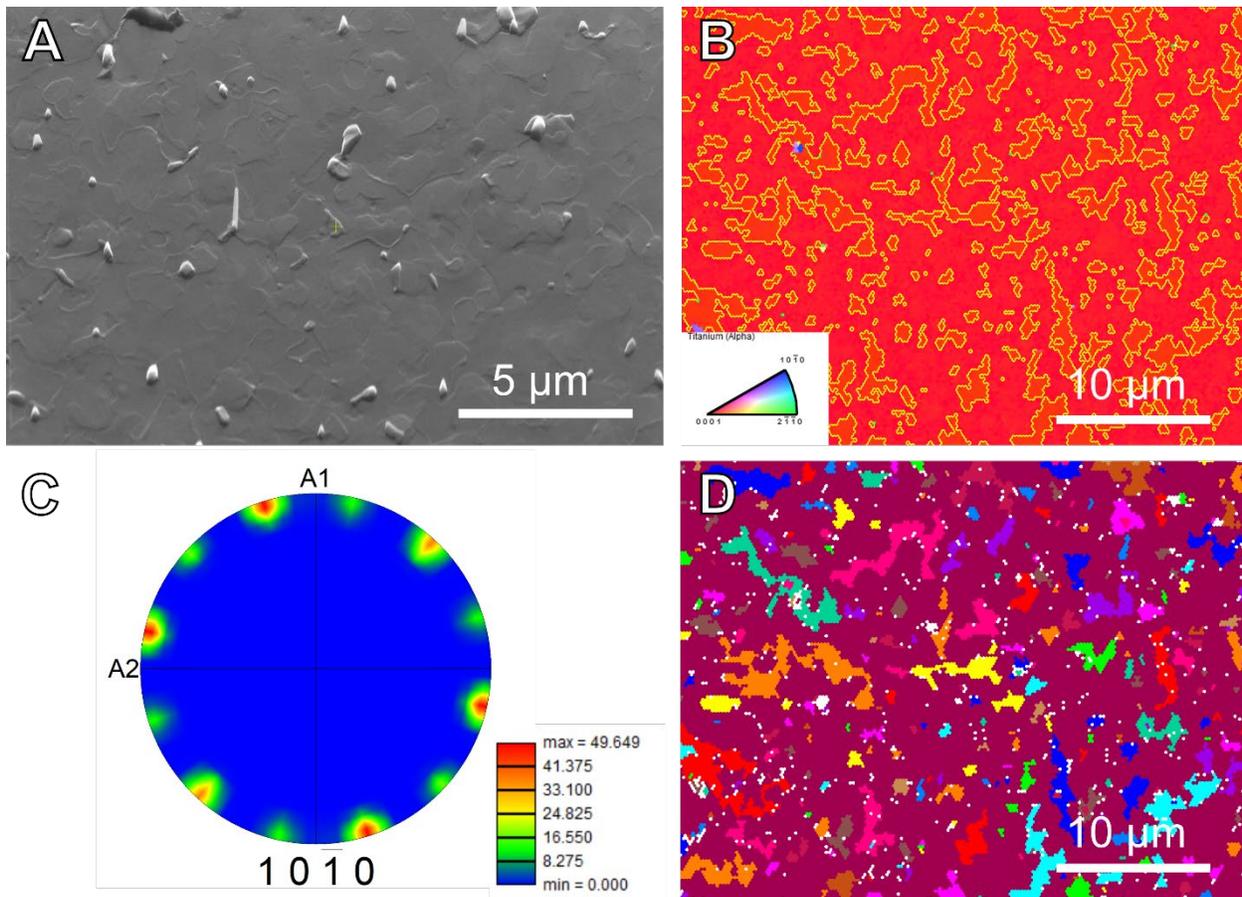

**Fig. S1. Ti thin film deposited using high power pulsed magnetron sputtering.** (**A**) Secondary electron image showing the roughly smooth surface and micron sized grains. (**B**) Orientation map obtained using electron backscatter diffraction (EBSD). The out-of-plane [0001] direction is indicated by red color and the Σ13 GBs are highlighted in yellow. (**C**) Pole figure obtained from the EBSD indicating a very strong [0001] texture and the bicrystalline nature of the film. (**D**) Unique grain map indicating that one of the grains extends all over the thin film while the grains belonging to second orientation are restricted as islands.

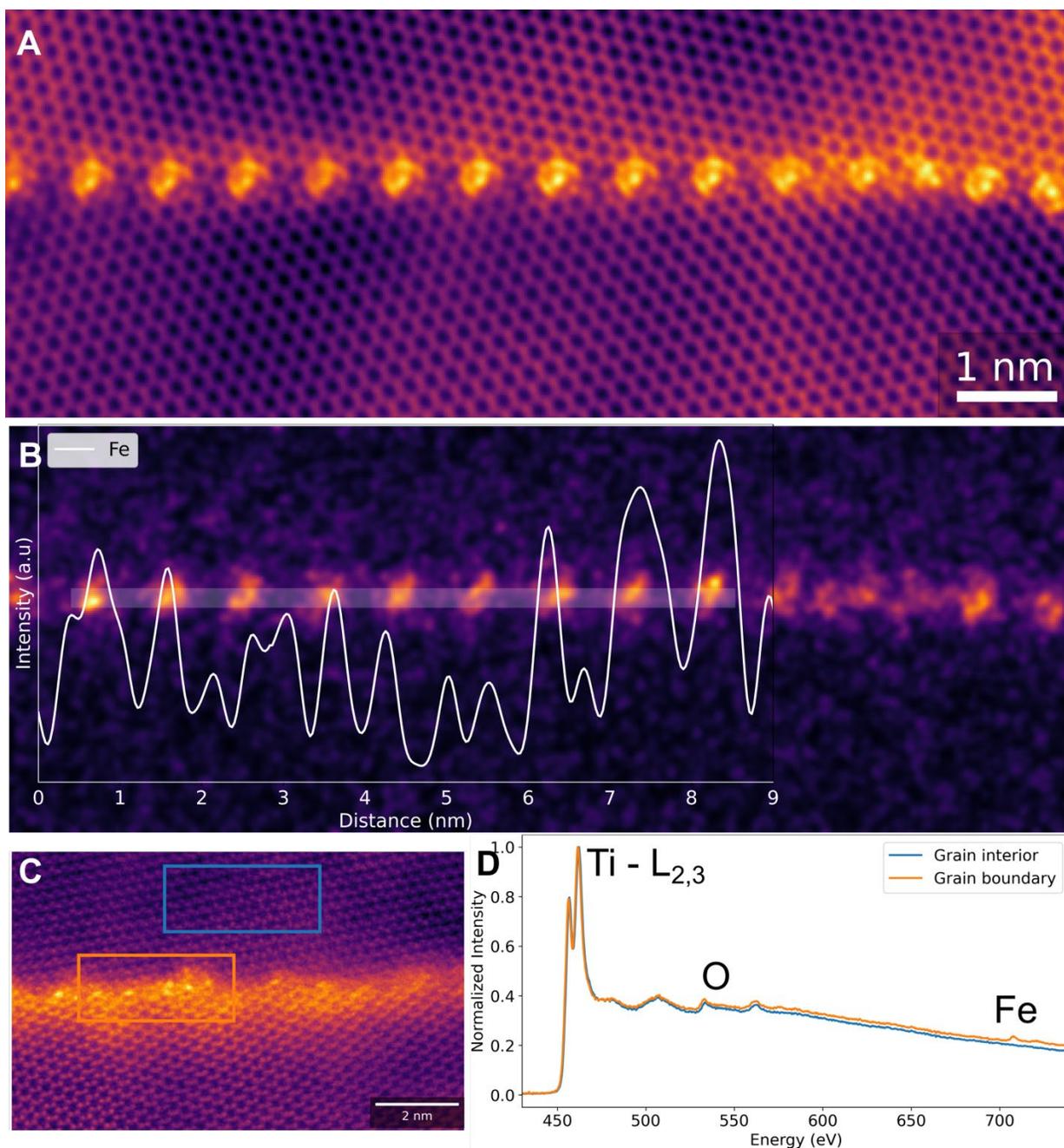

**Fig. S2. Energy-dispersive X-ray spectroscopy and electron energy loss spectroscopy of Fe-cages.** (**A**) HAADF-STEM image of the cages uniformly spaced along the grain boundary (**B**) STEM-EDS elemental map of Fe acquired from the corresponding area showing intense peaks corresponding to the cage centers. The 1-D line profile acquired along the GB (marked in white) highlights the Fe enrichment at periodic intervals corresponding to the cage centers. (**C**) HAADF-STEM image of the region utilized for electron energy loss spectroscopy (EELS), with the bulk grain interior and the GB region highlighted in blue and orange, respectively. (**D**) The corresponding EELS data shows a slight oxygen (O) peak that is nearly the same for the bulk grain (blue) and the GB (orange) and may originate from surface oxidation of the TEM specimen. The Ti-$L_{2,3}$ peak ratio remains very similar in the GB and within the grain.

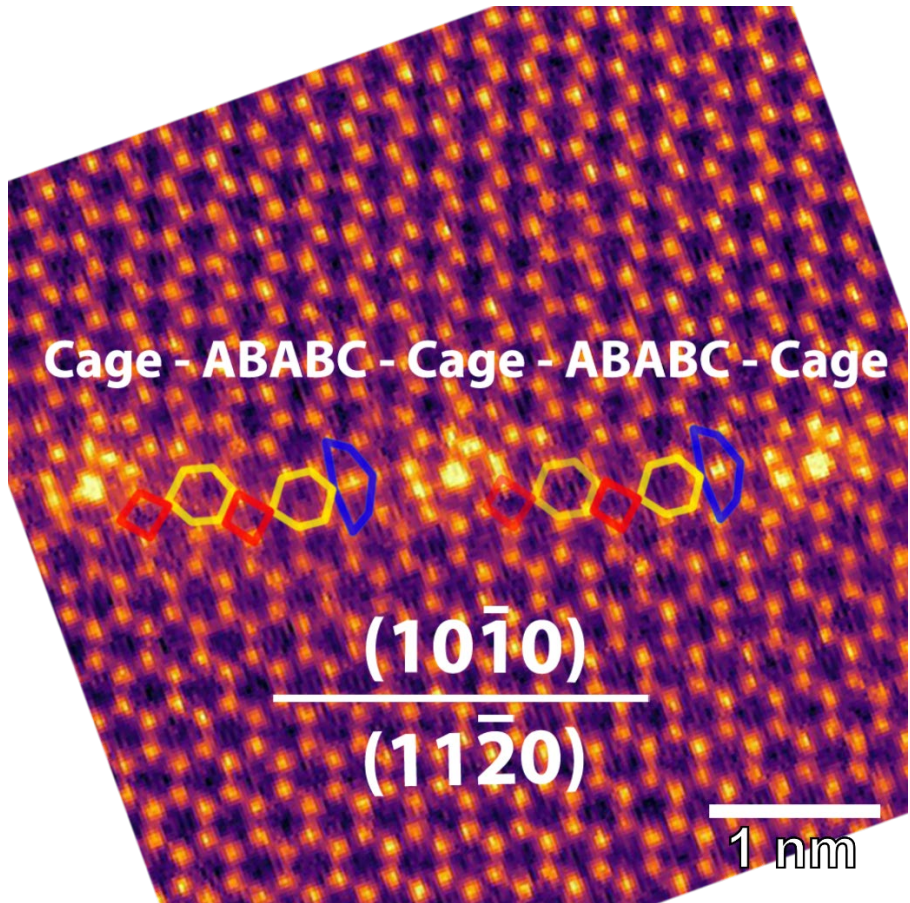

**Fig. S3. Coexistence of GB phases.** The $(10\bar{1}0)/(11\bar{2}0)$ asymmetric GB consists of a periodic arrangement of 'ABC' units that constitute the 'clean' GB and the cages.

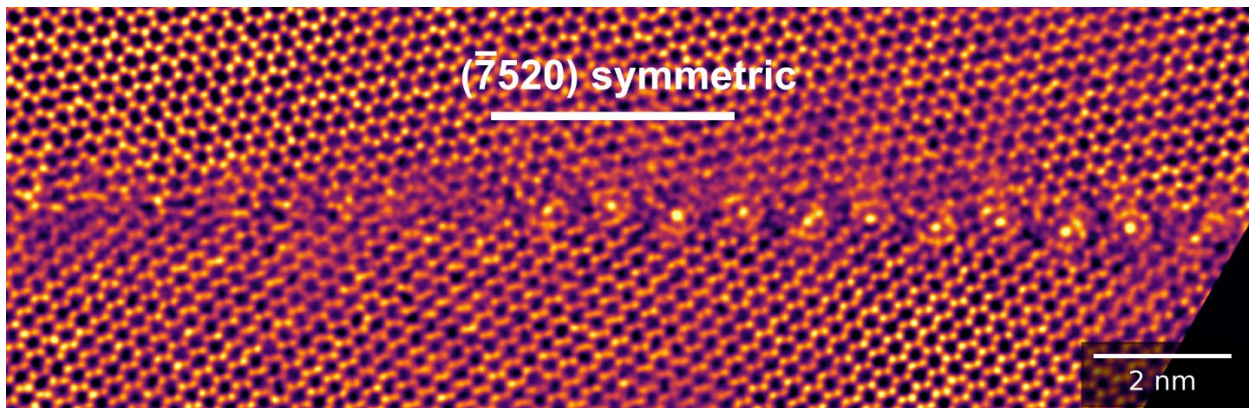

**Fig. S4. Coexistence of the initial GB ('ABC') with 'single-cage' GB phase on the same $\{\bar{7}520\}$ GB plane.

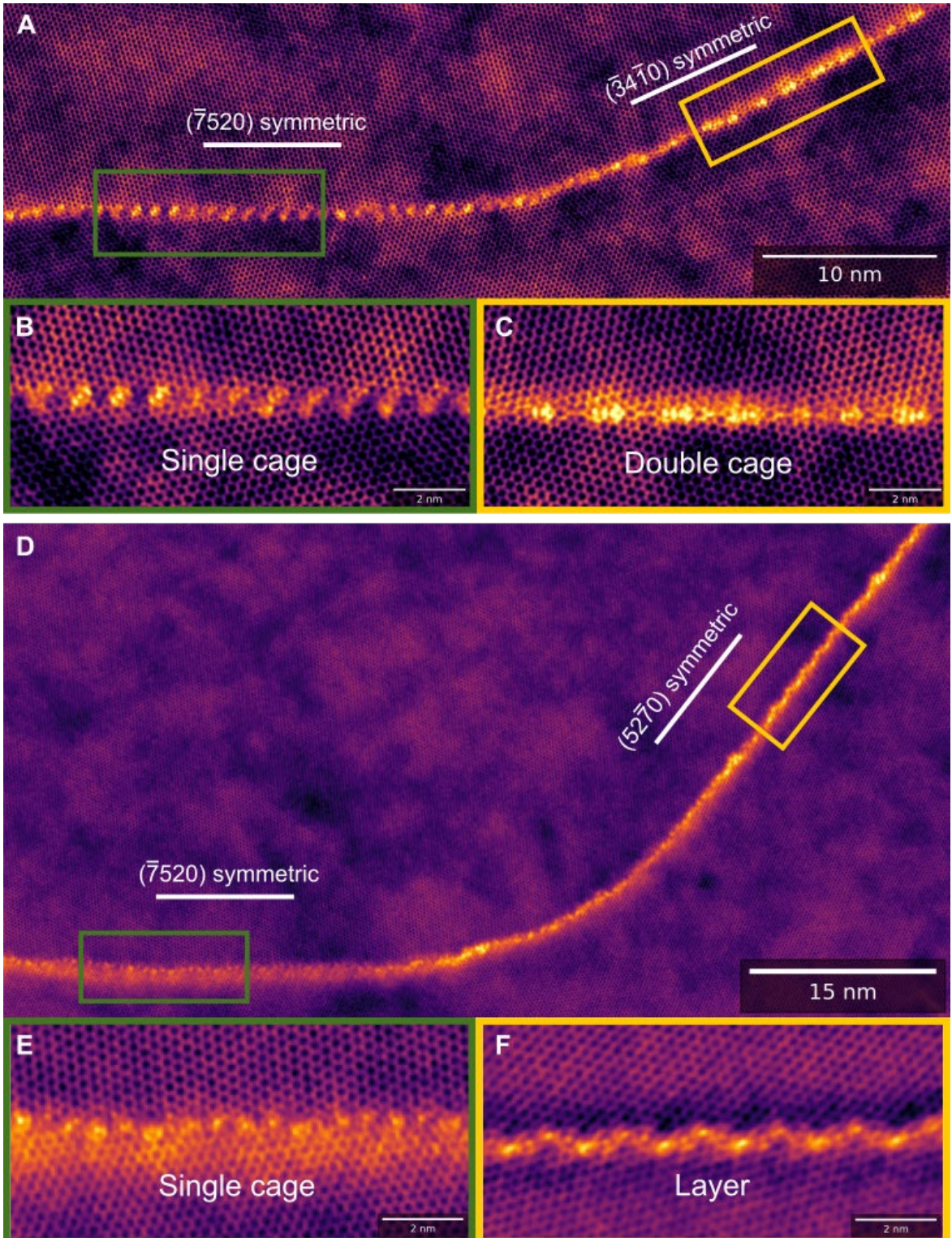

**Fig. S5. Segment of a curved GB containing single-cage, double-cage and layer-cage GB states.** (**A, D**) Atomically resolved STEM-HAADF images of Σ13 {$\bar{7}520$} [0001] GB with a large field of view. (**B, C**) Single-cage and double-cage structure is observed in the ($\bar{7}520$) and ($\bar{3}4\bar{1}0$) symmetric grain boundaries, respectively. (**E, F**) Single-cage and layer-cage structures are observed within the same symmetric GB plane family, ($\bar{7}520$) and ($52\bar{7}0$).

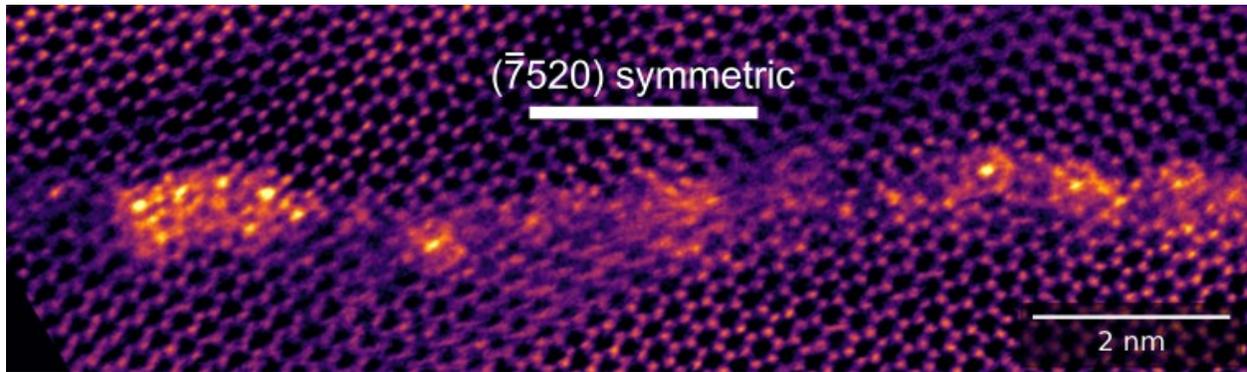

**Fig. S6. The largest isolated cluster of cages observed on a GB.** With a locally increased Fe concentration at the symmetric Σ13 {$\bar{7}520$} [0001] GB, the cages tend to form clusters as seen in the left of the HAADF-STEM image.

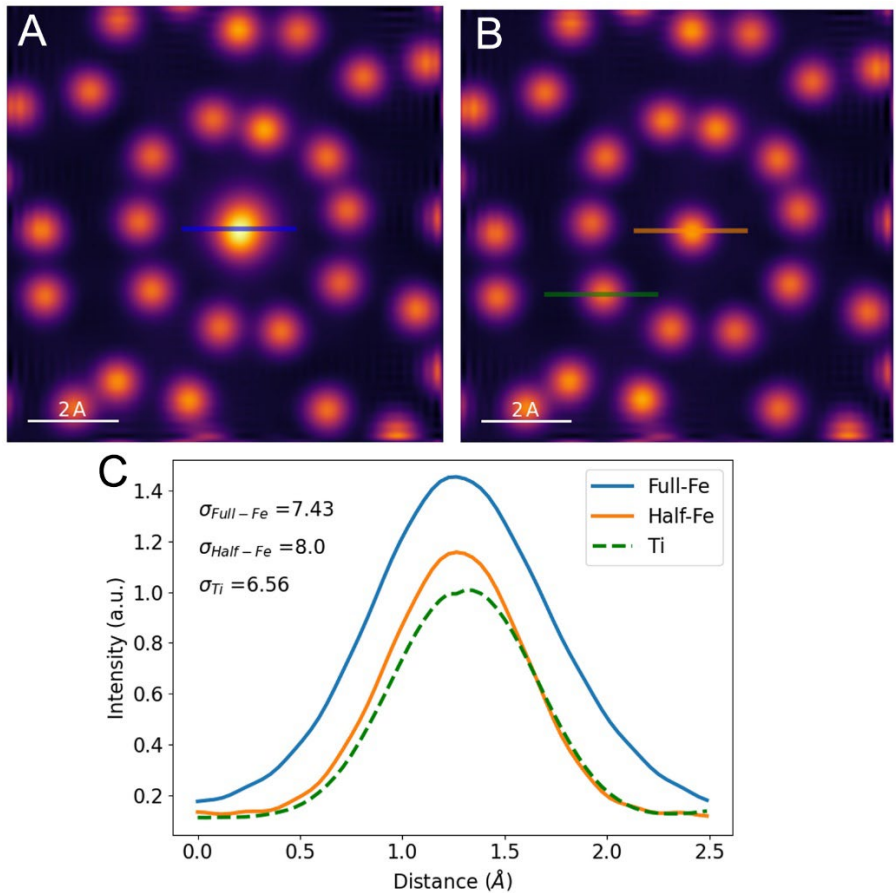

**Fig. S7. HAADF-STEM image simulations of the Fe-cages at the grain boundary.** (**A**) Central Fe-column that is completely filled. (**B**) Central Fe-column is only half filled with Fe, (**C**) Intensity of Fe-column in a), b) and Ti column in b) showing higher intensity and broader distribution of fully filled Fe-column in a).

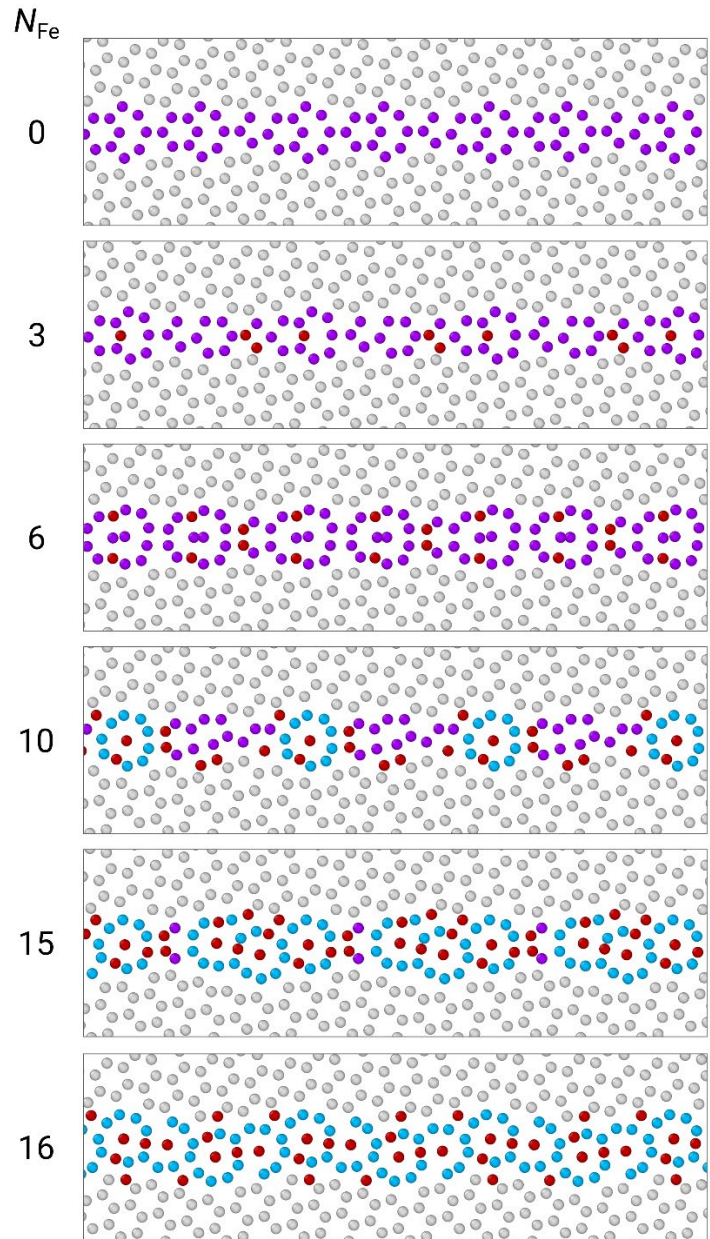

**Fig. S8. Grand canonically optimized Ti-Fe GBs using GRIP.** $N_{Fe}$ denotes the number of Fe atoms (red) per unit of the smallest periodic cell of the $\Sigma 13\{\bar{7}520\}[0001]$ GB. Cages are formed with increased Fe concentration (segregation). Ti atoms belonging to icosahedral cages are highlighted in blue and other GB Ti atoms are in purple. Bulk hcp Ti is in gray.